\documentclass[10pt,conference]{IEEEtran}
\usepackage{epsfig}
\usepackage{amsmath}
\usepackage{amsfonts}
\usepackage{amssymb}
\usepackage{calc}
\usepackage{xcolor}
\newcommand{\vsrem}[1]{{\color{lightgray} #1}}                   
\newcommand{\vscom}[1]{\,~\\{\footnotesize\color{red} #1}\,~\\}  
\newcommand{\vsadd}[1]{{\color{blue} #1}}                        
\newcommand{\remove}[1]{\color{.}}   
\newcommand{\comment}[1]{\color{.}}  
\newcommand{\add}[1]{#1}  
\newcommand{\vsaccept}{
    \let\vsrem\remove  %
    \let\vscom\comment %
    \let\vsadd\add     %
}%

\newcommand{\X}{{\mathbb X}}
\newcommand{\ext}{\mbox{\scriptsize ext}}
\newcommand{\dmax}{{d_{\mbox{\scriptsize max}}}}
\newcommand{\pth}{{p_{\mbox{\scriptsize th}}}}
\newcommand{\fp}{{\phi_p}}
\newcommand{\upth}{{\mbox{\scriptsize th}}}
\newcommand{\iter}{\delta}
\newlength{\mytablength}
\newcommand{\mytab}[1][1]{\noindent\hspace{#1\mytablength}}

\title{\LARGE Irregular turbo code design for the binary erasure channel}
\author{Ghassan M. Kraidy, Valentin Savin\\
CEA-LETI, 17 rue des Martyrs, 38054 Grenoble, France\\
\{ghassan.kraidy,valentin.savin\}@cea.fr}
\date{}
\begin{document}
\maketitle \vspace{-15mm}

\vsaccept 
\begin{abstract}
In this paper, the design of irregular turbo codes for the binary
erasure channel is investigated. An analytic expression of the
erasure probability of punctured recursive systematic convolutional
codes is derived. This exact expression will be used to track the
density evolution of turbo codes over the erasure channel, that will
allow for the design of capacity-approaching irregular turbo codes.
Next, we propose a graph-optimal interleaver for irregular turbo
codes. Simulation results for different coding rates is shown at the
end.
\end{abstract}


\section{Introduction}
The performance of error correcting codes over the binary erasure
channel (BEC) can be analyzed precisely, and a flurry of research
papers have already addressed this issue.  For small to medium
codeword length, {\em Maximum-Distance Separable} (MDS) codes
achieve the capacity of the BEC. However, for large block lengths,
their decoding becomes untractable, and thus iteratively decoded
graph-based codes present the main alternative. Low-density
parity-check (LDPC) codes \cite{di2002fla}
\cite{richardson_design_2001} \cite{luby2001eec} and
repeat-accumulate (RA) codes \cite{pfister2005cae} with
message-passing decoding proved to perform very close to the channel
capacity with reasonable complexity. Moreover, ``rateless'' codes
\cite{luby2002lc} \cite{shokrollahi2006rc} that are capable of
generating an infinite sequence of parity symbols were proposed for
the BEC. However, convolutional-based codes, that are widely used
for Gaussian channels, are less investigated on the BEC.  Among the
few papers that deal with convolutional and turbo codes over the BEC
are \cite{kurkoski_exact_2003} \cite{kurkoski_analysis_2004}
\cite{rosnes_turbo_2007} \cite{jeong_w_lee_performance_2007}. In
this paper, we propose irregular turbo codes that approach the
capacity of the BEC for medium to large block length. This is
accomplished through precise asymptotic analysis of the codes
together with a graph-optimal interleaver. The paper is organized as
follows: in Section \ref{irregular} we describe the model of the
irregular turbo code. Section \ref{exact} gives the exact erasure
probability at the output of a punctured RSC code. The asymptotic
design of irregular turbo codes is then discussed in Section
\ref{design}, while Section \ref{peg_interleaver} presents an
optimal graph-based interleaver for such codes. Section \ref{sim}
shows the performance of these codes and Section \ref{conc} gives
the concluding remarks.

\section{Irregular turbo codes}
\label{irregular} 
\begin{figure}[h]
\begin{center}
   \includegraphics[width=0.15\columnwidth,angle=-90]{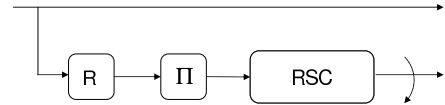}
   \vspace{4mm}
   \caption{Self-concatenated turbo encoder.}
   \label{tc_irreg}
\end{center}
\end{figure}

\begin{figure}[t]
\begin{center}
   \includegraphics[width=0.65\columnwidth,angle=-90]{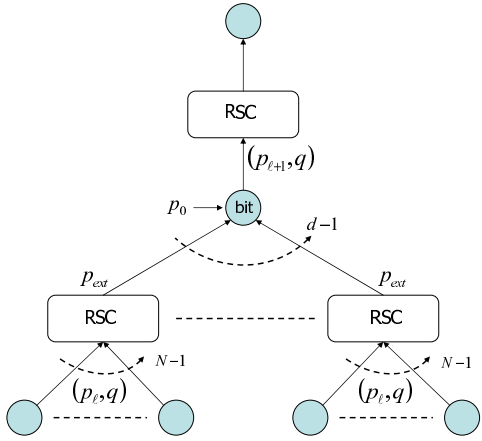}
   \vspace{4mm}
   \caption{Propagation tree of an irregular turbo code}
   \label{tree}
\end{center}
\end{figure}

\normalsize A parallel turbo code
\cite{berrou_near_1996} generally consists of a concatenation of two
recursive systematic convolutional (RSC) codes.  An information
sequence {\bf b} is encoded by the first RSC code to generate a
first parity bit sequence; the  same sequence is then scrambled by
an interleaver ${ \Pi}$ and encoded by a second RSC code to generate
a second parity bit sequence. In most cases, the two constituent RSC
encoders of a parallel turbo code are identical. For this reason,
the authors in \cite{boutros2001abs} \cite{boutros:tcd} proposed a
``self-concatenated'' turbo encoder in which every information bit
is repeated twice, interleaved, and fed to an RSC code of double the
size, as shown in Fig.~\ref{tc_irreg}.
\vsrem{In this new representation, each information bit is connected
to the code trellis via two edges in the propagation tree, as shown
in Fig.~\ref{tree}. We hence say that the {\em degree} of the
information bits is $d=2$, and that the turbo code is regular.}
 In this new representation, each information bit is
connected to the code trellis via two edges in the propagation tree
of Fig.~\ref{tree}. Therefore, we say that the {\em degree} of the
information bits is $d=2$, and that the turbo code is regular.
Using this structure, one can create irregularity by repeating a
certain fraction \vsrem{$f_i$} \vsadd{$f_d$} of the bits \vsrem{$i$}
\vsadd{$d$} times, providing bits that are more protected than in
the regular case. Like for LDPC and RA codes
\cite{richardson_design_2001} \cite{luby2001ild}, irregularity can
boost the performance of turbo codes for large block lengths.
Irregular turbo codes were first introduced in \cite{frey:itc}. In
\cite{boutros2001abs} \cite{boutros:tcd}, in a slightly different
design, a fraction of the information bits is repeated $d$ times
with $d>2$, while the parity bits remained of degree 1. In order to
maintain the same coding rate, a fraction \vsrem{$\phi_p$}
\vsadd{$\fp$} of the parity bits is punctured. We will use this
representation to design irregular turbo codes for the binary
erasure channel. The encoder of an irregular turbo code is similar
to that of Fig.~\ref{tc_irreg}, with the difference that the
repetition is non-uniform. The information bits are thus divided
into $d$ classes with $d=2,\dots,\dmax$, where \vsrem{$d_{max}$}
\vsadd{$\dmax$} is the maximum bit-node degree. The number of bits
in a class $d$ is a fraction $f_d$ of the total number of
information bits at the turbo encoder input, where bits in class $d$
are repeated $d$ times. Finally, the output of the non-uniform
repeater is interleaved and fed to the RSC constituent code, of
which $(1-\fp)$ of the parity bits are transmitted. Now let $K$
denote the length of the information sequence, $N$ the interleaver
size, \vsadd{$\rho_0$ and} $\rho$ \vsadd{the initial} and the final
(punctured) rate of the RSC constituent code respectively, and $R_c$
the rate of the turbo code. We can write the following:
\vspace{-4mm}

 \footnotesize
\begin{equation}
\sum_{d = 2}^{d_{max}} f_d = 1,~~
\sum_{d = 2}^{d_{max}} d . f_d = \overline{d},~~
N = K\sum_{d = 2}^{d_{max}} d . f_d = K . \overline{d}
\end{equation}
\begin{equation}
\label{rate} R_c = \frac{K}{K + \frac{N}{\rho} - N} = \frac{1}{1 +
\left( \frac{1}{\rho} - 1\right) \overline{d}}
\end{equation}
\begin{equation}
\label{rho} \rho = \frac{1}{1 + \left(1 - \fp
\right)\left(\frac{1}{\rho_0} - 1 \right)}
\end{equation}
\normalsize
 \vsrem{where $\rho_0 = k/n$ is the initial rate of the
constituent RSC code before puncturing. Using the above equations,
one can find, for a fixed target rate $R_c$, a good degree profile
${\bf f} = \{ f_2, f_3, ... , f_{d_{max}} \}$ that can boost the
performance of the code.}
\vscom{The ``good profile'' is not found from the above equations, but from the density evolution (next section).}
\vsadd{For a degree profile $\{ f_2, f_3, ... , f_{\dmax} \}$ and
using the above equations, one can compute the puncturing fraction
$\fp$  corresponding to a target rate $R_c$ . The performance of an
irregular turbo code will strongly depend on the degree profile and
the puncturing fraction or, more specifically, on the corresponding
puncturing pattern. In the following sections we will consider the
design of capacity-approaching irregular turbo codes over the BEC.
To do so, we will first compute the analytic expression of the
extrinsic erasure probability at the output of the punctured RSC
decoder that represents the key tool for the density evolution of
irregular turbo codes.}
\vsrem{In the following sections, we will consider the design of
such codes for the BEC, in a goal to approach the capacity.  To do
so, we will compute the analytic expression for the extrinsic
erasure probability at the output of the RSC decoder with
puncturing, expression that consists the basis of our design
algorithm.}

\section{Erasure probability of punctured RSC codes}
\label{exact}
 In this section, we will derive the exact erasure
probability of binary RSC codes, taking into account the puncturing
of parity bits. To do so, we will follow the steps of the method
proposed in \cite{kurkoski_exact_2003} used to compute the erasure
probability at the output unpunctured RSC codes. For the sake of
simplicity, we only consider half-rate codes with constraint length
$L= \nu + 1$, where $\nu$ is the memory of the code.
The same method applies to RSC codes with different rates.\\
We consider the following communication scheme: a uniformly
distributed sequence of bits {\bf b} of length $K$ is fed to a
binary RSC encoder that generates a sequence {\bf c} of parity bits
of length $N(1-\phi_p)$. During transmission\footnote{We consider
different erasure probabilities on information and parity bits, in
order to be able to distinguish between the extrinsic (corresponding
to information bits) and the communication (corresponding to parity
bits) channels for the density evolution computation}, a bit $b_i$
(respectively $c_j$) is either erased with probability $p$
(respectively $q$), or perfectly received with probability $1-p$
(respectively $1-q$). Let ${\bf b'}$ and ${\bf c'}$ be the received
sequences at the decoder. An RSC code has $S=2^{\nu}$ states.
Considering the ``Forward-Backward'' \cite{bcjr} decoding algorithm,
let $F_n(s)$ and $B_n(s)$ be the probabilities of being in state
$s=1,\dots, S$ computed in the forward and in the backward
directions, at the left and at the right side of the
$n^{\mbox{\scriptsize th}}$ trellis step respectively. Let $l(e)$
and $r(e)$ be the states to which an edge $e$ is connected on the
left and on the right respectively. The information bit $b(e)$ and
parity bit $c(e)$ are associated to edge $e$. As shown in
\cite{kurkoski_exact_2003}, the extrinsic probability of an
information bit $b_n$ at the output of the decoder is written
as:\vspace{-2mm}

\footnotesize
\begin{eqnarray}
P_{\rm ext}(b_n)  &=& P\left( b_n|{\bf b'}_{- \infty}^{n-1}, {\bf
b'}_{n+1}^{\infty},{\bf c'}\right) \nonumber \\ &\propto&
\sum_{e\,:\,b(e)=b_n} F_{n}(l(e))\cdot P(c(e))\cdot B_{n}(r(e))
\end{eqnarray}
\normalsize
Now the let $\Sigma_F = \{ \sigma_f^1, \dots, \sigma_f^{| \Sigma_F
|} \}$ and $\Sigma_B = \{ \sigma_b^1, \dots, \sigma_b^{| \Sigma_B |}
\}$ be the sets from which $F_n$ and $B_n$ take values. The
cardinality of the sets $\Sigma_F$  and $\Sigma_B$ is computed
as:\vspace{-2mm}

\footnotesize
\begin{equation}
| \Sigma_F | = | \Sigma_B | = \sum_{\alpha=0}^{\nu}
\binom{2^{\alpha}}{2^{\nu}}
\end{equation}
\normalsize \vsrem{where the entries are normalized to 1.} However,
as an RSC code is linear, we assume the all-zeros codeword is
transmitted without losing generality. This gives smaller state
\vsrem{metric} \vsadd{distribution} sets $\Sigma_F^*$ and
$\Sigma_B^*$ with cardinality:

\footnotesize
\begin{equation}
| \Sigma_F^* | = | \Sigma_B^* | = \sum_{\alpha=0}^{\nu}
\binom{2^{\alpha}-1}{2^{\nu}-1}
\end{equation}
\normalsize A four-state RSC code ($L=3$) has for
instance:\vspace{-2mm}

\footnotesize
\begin{eqnarray}
 \Sigma_F^* = \Sigma_B^* &=&
 \{ (1,0,0,0),(1/2,1/2,0,0),(1/2,0,1/2,0), \nonumber\\ & & (1/2,0,0,1/2),(1/4,1/4,1/4,1/4) \}
\end{eqnarray}
\normalsize

\subsection{Computation of the Erasure Probability}
\vscom{I have introduced forward and backward matrices
$M_{F,\X}(p,q)$ and $M_{B,\X}(p,q)$, depending on the puncturing
pattern $\X$. These matrices are computed from previously introduced
matrices $M_{F}(p,q)$ and $M_{B}(p,q)$. There are also some other
minor changes (mainly concerning the notation)} The trellis of a
convolutional code forms two first-order $S$-state Markov chains
corresponding to the forward and backward recursions. This allows to
compute the steady-state distributions of the Markov processes that
will be used to compute the bit erasure probability at the output of
the decoder. The distributions $\pi_F(p,q)$ and $\pi_B(p,q)$ are the
\vsadd{normalized} solutions \vsrem{to} \vsadd{of} the following
equations:\vspace{-3mm}

\footnotesize
\begin{equation}
\pi_F(p,q) =\pi_F(p,q)\cdot M_F(p,q);~~ \pi_B(p,q) = \pi_B(p,q)\cdot
M_B(p,q)
\end{equation}
\normalsize
 where the $(i,j)^{\upth}$ entry of matrix
$M_F$ is the probability of the transition from state distribution
$F_n = \sigma_f^i$ to state distribution $F_{n+1} = \sigma_f^j$.
Similarly, the matrix $M_B$ represents the transition probabilities
in the backward direction.  In other words, the distributions
$\pi_F$ and $\pi_B$ are the stationary distributions to which the
Markovian process converges, as:

\footnotesize
\begin{equation}
\lim_{\iter \rightarrow \infty} M_F^{\iter} = {\bf 1} \otimes
\pi_F;~~~~ \lim_{\iter \rightarrow \infty} M_B^{\iter} = {\bf 1}
\otimes
 \pi_B
\end{equation}
\normalsize

\noindent where {\bf 1} is a column vector of ones. As an example,
we will consider the four-state RSC $(1,5/7)_8$ code with $L=3$. Let
$p$ be the erasure probability on the information bits, and $q$ be
the erasure probability on parity bits. Assuming the all-zeros
codeword has been transmitted, we have that $| \Sigma_F^* | = |
\Sigma_B^* |=5$. The $5 \times 5$ Markov state transition matrix
$M_F$ for the forward recursion of this code is given
by:\vspace{-2mm}

\footnotesize
\begin{equation*}
\begin{array}{l}
{M}_{F}(p,q) =\\~\\  \left[\begin{array}{ccccc}
1 - pq & 0 & pq & 0 & 0\\
0 & 0 & 1 & 0 & 0\\
1 + pq - p - q & p - pq & 0 &
q - pq & pq\\
1 + pq - p - q & q - pq & 0 & p -pq & pq \\
0 & 0 & 1 + pq - p - q & 0 & p + q - pq
\end{array}\right]
\end{array}
\end{equation*}
\normalsize and the matrix $M_B$ for the backward recursion is given
by:\vspace{-2mm}

\footnotesize
\begin{equation*}
\begin{array}{l}
{M}_{B}(p,q) =\\~\\ \left[\begin{array}{ccccc}
1 - pq & pq & 0 & 0 & 0\\
1 + pq - p - q & 0 & p - pq & q -pq & pq \\
0 & 1 & 0 & 0 & 0\\
1 + pq - p - q & 0 & q -pq & p - pq & pq \\
0 & 1 + pq - p - q & 0 & 0 & p + q - pq
\end{array}\right]
\end{array}
\end{equation*}
\normalsize Once ${M}_{F}(p,q)$ and ${M}_{B}(p,q)$ are computed,  we
can solve for $\pi_F(p,q)$ and $\pi_B(p,q)$. We next consider the
matrix $T(q)$ whose $(i,j)^{\mbox{\scriptsize th}}$ entry represents
the probability of an output erasure conditioned on the left and
right state distributions $\sigma_f^i$ and $\sigma_b^j$, knowing
that parity bits are erased with probability $q$:

\footnotesize
\begin{equation}
T_{i,j}(q) = P \left(P_{\rm ext}(b_n)=1/2 \mid F_n=\sigma_f^i,
B_{n}=\sigma_b^j \right)
\end{equation}
\normalsize

The matrix $T(q)$ for the RSC $(1,5/7)_8$ code is given by:

\footnotesize
\begin{equation*}
\begin{array}{l}
{ T}(q) =\left[\begin{array}{ccccc}
0 & 0 & q & 0 & 0\\
q & q & q & q & q\\
0 & 1 & q & 0 & 1\\
0 & 0 & q & 1 & 1\\
q & 1 & q & 1 & 1
\end{array}\right]
\end{array}
\end{equation*}
\normalsize Finally the extrinsic erasure probability is computed
as:

\footnotesize
\begin{equation}
\label{ext} P_{\rm ext}(p,q) = \pi_F(p,q) \cdot T(q) \cdot
\pi_B(p,q)^t
\end{equation}
\normalsize where the operator $(.)^t$ denotes the transpose
operator.
\subsection{Computation of the Erasure Probability with puncturing}
\label{ext_proba_punct} Now suppose a fraction $\fp$ of the parity
bits of the code are punctured. If the punctured parity bits were
randomly chosen at each transmission, we could consider that the
fraction $\fp$ of punctured bits is a part of the channel, as if the
decoder receives bits with probability of erasure on parity bits
given by:\vspace{-2mm}

\footnotesize
\begin{equation}
\label{new_erasure} q' = 1 - \left( 1- q \right) \left( 1 - \fp
\right) = \fp + q - q\cdot \fp
\end{equation}
\normalsize

However, if the puncturing pattern is fixed, the extrinsic erasure
probability computed using (\ref{ext}) by replacing $q$ with $q'$
from (\ref{new_erasure}) is inaccurate. The goal is then to
analytically compute the extrinsic erasure probability at the output
of the decoder knowing that parity bits are punctured using a
predefined pattern. For this purpose, we define a puncturing pattern
$\X=[x_1,x_2, ..., x_\Gamma],~x_\gamma \in \{0,1\}$, where a $0$ in
position $\gamma$ means that the parity bit in the corresponding
trellis step is punctured. The parity bits of the constituent RSC
code are then punctured using a periodic puncturing pattern with
period $\X$. We consider a window of size $\Gamma$ in the trellis of
the code, and let $M_{F,\X}(p,q)$ the matrix whose $(i,j)^{\upth}$
entry is the probability of the transition from the state
distribution $F_n = \sigma_f^i$ at the left side of the window  to
the state distribution $F_{n+\Gamma} = \sigma_f^j$ at the right side
of the window. Similarly, the matrix $M_{B,\X}$ represents
``throughout-the-window'' transition probabilities in the backward
direction. We have the following:\vspace{-4mm}

\footnotesize
\begin{equation}
M_{F,\X}(p,q) = \prod_{\gamma = 1}^{\Gamma} M_{F}(p,q^{x_\gamma});~~
M_{B,\X}(p,q) = \prod_{\gamma = 1}^{\Gamma} M_{B}(p,q^{x_{\Gamma+1 -
\gamma}})
\end{equation}
\normalsize

This means that $M_{F,\X}(p,q)$ is obtained by multiplying matrices
$M_F(p,1)$ and $M_F(p,q)$ according to whether the corresponding
parity bit is punctured ($x_\gamma = 0$) or not ($x_\gamma = 1$). A
similar assertion holds for the backward matrix $M_{B,\X}(p,q)$.

Let $\pi_{F,\X}(p,q)$ and $\pi_{B,\X}(p,q)$ be the corresponding
steady-state distributions, meaning that:

\footnotesize
\begin{equation}
\lim_{\iter \rightarrow \infty} M_{F,\X}^{\iter} = {\bf 1} \otimes
\pi_{F,\X};~~ \lim_{\iter \rightarrow \infty} M_{B,\X}^{\iter} =
{\bf 1} \otimes
 \pi_{B,\X}
\end{equation}
\normalsize where {\bf 1} is a column vector of ones. These
expressions represent the state probability distributions in the
forward and backward directions, at the left and at the right side
of the window respectively. The distributions $\pi_{F,\gamma}(p,q)$
on the left side and $\pi_{B,\gamma}(p,q)$ on the right side of a
window step $\gamma$ can be recursively computed as:\vspace{-2mm}

\footnotesize
\begin{eqnarray}
\pi_{F,1}(p,q) &=& \pi_{F,\X}(p,q),\nonumber \\  \pi_{F,\gamma}(p,q) &=& \pi_{F,\gamma-1}(p,q) \cdot {M}_{F}(p,q^{x_\gamma}),~ \gamma = 2,\dots, \Gamma\\
\pi_{B,\Gamma}(p,q) &=& \pi_{B,\X}(p,q),\nonumber \\
\pi_{B,\gamma}(p,q) &=& \pi_{B,\gamma+1}(p,q) \cdot
{M}_{B}(p,q^{x_\gamma}),~ \gamma = \Gamma-1,\dots, 1
\end{eqnarray}
\normalsize
 Next, the extrinsic erasure probability of the
information bit in position $\gamma$ can be computed as:

\footnotesize
\begin{equation}
\label{ext_gen} P_{\ext,\gamma}(p,q) = \pi_{F,\gamma}(p,q) \cdot
T(q^{x_\gamma}) \cdot {\pi_{B,\gamma}(p,q)}^t
\end{equation}
\normalsize

\noindent Finally, the extrinsic erasure probability at the output
of the decoder corresponding to the puncturing pattern $\X$ is given
by:

\footnotesize
\begin{equation} \label{p_punct} P_{\ext,\X}(p,q) =
\frac{1}{\Gamma} \sum_{\gamma=0}^{\Gamma} P_{\ext,\gamma}(p,q)
\end{equation}
\normalsize

As an example, suppose we want to construct a half-rate parallel
turbo code using half-rate $(1,5/7)_8$ RSC codes.  In order to raise
the rate of the constituent codes from $1/2$ to $2/3$, we puncture
half of their parity bits using the pattern $\X=[1,0]$. The
expression of the exact probability of this code can then be written
as:

\footnotesize
\begin{equation}
P_{\ext,\X}(p,q) = \frac{1}{2} \left[P_{\ext,1}(p,q)+P_{\ext,2}(p,q)
\right]
\end{equation}
\normalsize

where\vspace{-2mm}

\footnotesize
\begin{eqnarray} P_{\ext,1} &=& \pi_{F,\X}(p,q)\cdot
T(q)\cdot \left[ \pi_{B,\X}(p,q)\cdot {M}_{B}(p,1)
\right]^t\\
P_{\ext,2} &=& \left[\pi_{F, \X}(p,q)\cdot{M}_{F}(p,q)\right]\cdot T(1) \cdot \pi_{B,\X}(p,q)^t
\end{eqnarray}
\begin{eqnarray}
\pi_{F,\X}(p,q) &=&\frac{1}{5}\cdot{\bf 1}^t\cdot \lim_{\iter \rightarrow \infty} M_{F,\X}^{\iter}(p,q),\mbox{ and } \nonumber\\
{M}_{F,\X}(p,q) &=& {M}_{F}(p,q)\cdot {M}_{F}(p,1) \\
\pi_{B,\X}(p,q) &=& \frac{1}{5}\cdot{\bf 1}^t\cdot \lim_{\iter
\rightarrow \infty} M_{B,\X}^{\iter}(p,q), \mbox{ and }\nonumber \\
{M}_{B,\X}(p,q) &=& {M}_{B}(p,1)\cdot {M}_{B}(p,q)
\end{eqnarray}
\normalsize

The exact expression of the erasure probability in (\ref{p_punct})
is the key tool for designing irregular turbo codes for the BEC, as
will be discussed in the following section. In fact, for the same
puncturing fraction $\fp$, it is capable of determining which
pattern $\X$ gives the lowest $P_{\ext,\X}$. Moreover, it allows to
detect a catastrophic puncturing scenario that leads to infinite
error events and thus harms the correction capacity of the code. As
an example, puncturing the RSC $(1,5/7)_8$ code using $\X=[1,0,0]$
gives:

\footnotesize
\begin{equation} P_{\ext,\X}(0,q) > 0
\end{equation}
\normalsize This means that this puncturing pattern is catastrophic,
as a single bit error at the input of the decoder generates an
infinite error event. If we have $\phi_p = 2/3$, we would rather use
$\X=[1,0,0,0,1,0]$ for instance. Although this example can be
directly observed on the trellis of the $(1,5/7)_8$ code,
(\ref{p_punct}) points out the phenomenon for any $S$-state code (on
any channel!), where trellis analysis becomes more tedious as $S$
increases.

\section{Irregular turbo code design}
\label{design}
 The analytic expression of the erasure probability of
punctured RSC codes in the previous section allows us to analyze the
iterative decoding of turbo codes over the BEC.  As discussed in
Section \ref{irregular}, a parallel turbo code consists of a
parallel concatenation of two RSC codes. The iterative decoding of
such codes can be analyzed through EXIT charts
\cite{tenbrink2001cbi}, that are non-linear functions relating the
output to the input of the RSC decoders of the infinite-length turbo
code. This technique gives insight on the iterative process in the
sense that the decoding is successful for a certain channel quality
if the two curves corresponding to the two decoders do not
intersect. The threshold of the code is the worst value of the
channel quality at which the tunnel between the two curves is open.
In the case where the two constituent codes are identical, the
decoding converges if the curve of the RSC decoder does not
intersect with the forty-five degree line.  Over the BEC, and with
the difference of Gaussian channels in general, an EXIT chart
describing the iterative decoding process gives the exact density
evolution of erasure probabilities, as we can compute analytic
expressions of the output as a function of the input of the decoder.
Although widely used for LDPC codes, this property was first
exploited in \cite{measson2003fap} to compute exact thresholds for
regular unpunctured turbo codes. For the sake of infinite-length
analysis, we represent a turbo code using the tree structure as
shown in Fig.~\ref{tree}, in which an information bit of degree $d$
is connected to $d$ trellises. For the regular parallel turbo code,
the erasure probability at iteration ${\ell + 1}$ is given by:

\footnotesize\begin{equation} \label{DE_equ}
 P_{\ell +1} = p_0 \cdot P_{\ext,\X} \left( P_{\ell},
p_0 \right)
\end{equation}\normalsize
where $p_0$ is the channel erasure probability. This expression
determines the density evolution of the iterative decoding process,
as it relates the probability at an iteration to that of the
previous iteration. Using (\ref{DE_equ}), the threshold probability
$\pth$ of the $R_c=1/3$ parallel turbo code built from rate-half RSC
$(1,5/7)_8$ constituent codes is computed as $\pth=0.6428$, knowing
that the capacity of the BEC is $C = 1 - p_0$. Again, the punctured
$R_c=1/2$ turbo code built from the same constituent codes has
$\pth=0.4729$.

In order to tighten the gap to the capacity of the BEC at a given
rate, we consider the design of irregular turbo codes.
\vsrem{For a bit of degree $d$, the probability of erasure after one iteration is
given by:}
\vsadd{The erasure probability at iteration $\ell + 1$ of a bit of
degree $d$ can be expressed as a function of the erasure probability
at iteration $\ell$ as:}

\footnotesize\begin{equation} P_{\ell +1}(d) = p_0 \cdot
\frac{d\cdot f_d}{\overline{d}}\cdot P_{\ext,\X} \left( P_{\ell},
p_0 \right)^{d-1}
\end{equation}\normalsize
\vsadd{Let $\lambda_d = \displaystyle\frac{d\cdot
f_d}{\overline{d}}$ and $\lambda(X) = \displaystyle
\sum_{d=1}^{\dmax} \lambda_d X^{d-1}$ (this definition will be made
clearer in Section \ref{peg_interleaver} where we will introduce the
factor graph of the turbo code). We can then write:

\footnotesize\begin{equation} P_{\ell +1}(d) = p_0 \cdot
\lambda_d\cdot P_{\ext,\X} \left( P_{\ell}, p_0 \right)^{d-1}
\end{equation}\normalsize
Averaging over all possible bit degrees, we get:

\footnotesize\begin{equation} \label{de_formula} P_{\ell +1} = p_0
\cdot \lambda\circ P_{\ext,\X}\left( P_{\ell}, p_0 \right)
\end{equation}\normalsize
Following this equation, the irregular turbo code can recover from a channel erasure probability $p_0$ if and only if

\footnotesize\begin{equation} p_0 \cdot \lambda\circ
P_{\ext,\X}\left( x, p_0 \right) \leq x,\ \forall x\in[0,p_0]
\end{equation}\normalsize
The code threshold is defined as:\vspace{-3mm}

\footnotesize\begin{equation} \pth(\lambda, \X) = \max\{p_0 \mid p_0
\cdot \lambda\circ P_{\ext,\X}\left( x, p_0 \right) \leq x,\ \forall
x\in[0,p_0]\}
\end{equation}\normalsize
and it depends on both degree distribution and puncturing pattern.
The design of capacity approaching irregular turbo codes reduces to
the optimization of the function $(\lambda,\X) \mapsto \pth(\lambda,
\X)$. For instance, this can be carried out using the differential
evolution algorithm \cite{storn1997des}. In general, a uniform
puncturing pattern leads to the best threshold, provided the pattern
is not catastrophic (which leads to a threshold equal to zero!).
Therefore, in order to reduce the space of parameters of the
optimization function, for each degree distribution $\lambda$, we
compute the puncturing fraction $\fp$ according to the target rate
$R_c$, and chose the puncturing pattern $\X$ as uniform as possible
according to $\fp$.} As an example, using the half-rate RSC
$(1,5/7)_8$ code and by setting $d_{\max}=12$, we obtained the
degree profiles in Table \ref{table_1}.

\begin{table}[h!]
 \caption{Degree profile of irregular turbo codes
over the BEC}\vspace{-4mm} \label{table_1} \scriptsize
$$\begin{array}{|*{11}{@{\:}c@{\:}|}}
\hline R_c      & f_2 & f_4 & f_5 & f_7 & f_8 & f_9 & f_{12} & \overline{d} & \phi_p & p_{th}\\
\hline
1/2        &  0.801 & 0.101 & & & 0.046 & & 0.052& 2.998            & 0.666 & 0.490\\
\hline
1/3        &  0.838 & & 0.034 & 0.041& 0.042 & 0.045 & & 2.873            & 0.304 & 0.665\\
\hline
1/4        &  0.837 & & 0.055 & & 0.054 & & 0.054    &   3.033           & 0.011 & 0.743\\
\hline

\end{array}$$
\label{convergence_1}
\end{table}
\vspace{-4mm}

\normalsize Note that the half-rate irregular turbo code designed
through differential evolution has $2$ parity bits out of $3$
punctured, and the optimization algorithm avoided catastrophic
puncturing while computing (\ref{p_punct}), as explained at the end
of Section \ref{exact}.

\section{PEG-based interleaver for turbo codes}
\label{peg_interleaver} 
\begin{figure}[h]
\begin{center}
   \includegraphics[width=0.25\columnwidth,angle=-90]{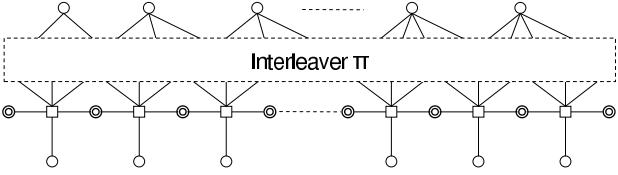}
   \vspace{4mm}
   \caption{Factor graph of irregular turbo codes.}
   \label{turbo_factor_graph}
\end{center}
\end{figure}

We investigate now the design of graph-based
interleavers for irregular turbo codes based on the progressive-edge
growth (PEG) algorithm \cite{PEG}. To do so, we define the factor
graph of an irregular turbo code, in a  manner similar to that of
\cite{wiberg}, \cite{vontobel2002ctc}. As shown in
Fig.~\ref{turbo_factor_graph} the factor graph consists of:
\begin{itemize}
\item bit nodes, represented by simple circles (information bits are represented on the top, while parity bits are represented on the bottom);
\item state nodes, represented by double circles;
\item trellis step nodes, also called transition nodes, represented by squares.
\end{itemize}
If $\rho_0 = k/n$ is the rate of the constituent RSC codes, then
each transition node is connected to $k$ information bits and $n-k$
parity bits. Using this representation, the previously defined
$\lambda_d = \displaystyle\frac{d\cdot f_d}{\overline{d}}$ is equal
to the fraction of edges emanating from information bit nodes of
degree $d$, and $\lambda = (\lambda_2,\dots,\lambda_{\dmax})$ is
called the {\em edge perspective} degree distribution.
\vspace{-4mm}

The decoding of turbo codes can be performed on the factor graph, by
iteratively propagating extrinsic messages from each graph node to
its neighbor nodes. As discussed in
\cite{boutros1997cat}\cite{vontobel2002ctc}
\cite{breiling_logarithmic_2004}, small cycles must be avoided in
the factor graph of a turbo code so that it looks locally tree-like,
and thus the messages are more independent. An upper-bound on the
girth (minimal cycle length) of an irregular factor graph can be
derived using a straightforward variation of the approach in
\cite{galager_monograph} (see also Lemma 2 in
\cite{PEG}):\vspace{-1mm}

\footnotesize
\begin{equation}
\label{g_upper_bound} g\leq 4\left( \left\lfloor
\frac{\log\left[(N-1)\frac{k}{k+2} +1
\right]}{\log(k+1)}\right\rfloor +1  \right)
\end{equation}
\normalsize Thus, a graph-optimal interleaving algorithm for
irregular turbo codes would yield factor graphs with girths that
grow as the logarithm of the interleaver size. Graphs with large
girths have been already used for the construction of regular turbo
codes \cite{vontobel2002ctc} and LDPC codes \cite{PEG}. The {\em
Progressive Edge Growth} algorithm proposed in \cite{PEG} is based
on a simple but very efficient idea: it progressively establishes
``best-effort'' connections in the graph, where a best-effort
connection corresponds to an edge maximizing the graph girth. In
what follows, we extend this construction to the case of irregular
turbo codes. The corresponding interleaver will be called PEG
interleaver.

The algorithm is submitted with the set ${\cal B} =
\{b_1,\cdots,b_K\}$ of information bit nodes, the set ${\cal T} =
\{t_1,\cdots,t_N\}$ of transition nodes, and a desired information
bit degree distribution. According to the submitted distribution, we
can write ${\cal B}$ as a disjoint union
 ${\cal B} = \cup_{d=2}^{\dmax} {\cal B}_d$,
where ${\cal B}_d$ is be the set of information bits with submitted
degree $d$. The algorithm starts with a factor graph comprising the
set ${\cal B}$ of information bit nodes, the set ${\cal T}$ of
transition nodes and the corresponding set of state nodes, each
transition node being connected to its left and right state nodes.
At this moment there is no connection between information bit and
transition nodes. We then progressively add edges emanating from
bits in the set ${\cal B}_2$, until all these bits reach the
submitted degree $2$. We next progressively connect the bits from
the sets ${\cal B}_3, \dots, {\cal B}_\dmax$. It is important to
notice that no bit of ${\cal B}_d$ is connected, as long as there
are bits in ${\cal B}_{d-1}$ that do not reach the submitted degree
($d-1$). This is done in order to protect the bits of small degree
in the following sens: when information bits of small degree ({\em
e.g.} $d = 2$)  are connected, the graph girth is relatively large;
this will help to avoid short cycles that contain only information
bit nodes of small degree. Adding more edges in the graph, the girth
will decrease. When information bit nodes of higher degree are
connected, we get smaller cycles, but these cycles are better
connected to other cycles in the graph. To connect bit nodes in
${\cal B}_d$ we proceed as follows: \vspace{2mm}

\footnotesize
 {\bf Progressive Edge Growth for ${\cal B}_d$}

\noindent{\bf for} $i=1,\dots,d$

    \mytab{\bf for} each $b\in{\cal B}_d$

        \mytab[2]{\bf if} $i = 1$

            \mytab[3] \begin{minipage}{\linewidth - 3\mytablength}
                      Choose a transition node $t$ of lowest degree in the current graph and connect $b$ to $t$
                      \end{minipage}

        \mytab[2]{\bf else}

            \mytab[3] \begin{minipage}{\linewidth - 3\mytablength}
                      Expand the current graph as a tree rooted at $b$, until all transition nodes are in the tree. Identify transition nodes that are connected to at most $k-1$ information bits in the current graph. Among these transition nodes, identify  those of maximal depth in the tree. Among these last identified transition nodes, choose a transition node $t$ of lowest degree in the current graph and connect $b$ to $t$
                      \end{minipage}

        \mytab[2]{\bf end}

    \mytab{\bf end}

\noindent{\bf end} \normalsize \vspace{1mm}

\noindent Note that, we first add an edge for each bit in ${\cal
B}_d$, then a second edge for each bit, and so on, until all  bits
reach the degree $d$.

In case that the puncturing pattern is known at the time of the
interleaver construction, we can use a slightly modified version of
the above algorithm as follows: first, each bit $b\in{\cal B}$ is
connected to an unpunctured transition node of lowest degree in the
current graph (we say that a transition node is unpunctured if its
parity bit is unpunctured). In case that the number of unpunctured
transition nodes is less than the number of information bit nodes,
we connect the information bits of smallest submitted degree.
Finally, the interleaver is constructed by running the above PEG
algorithm, but starting from this graph. In this way, we make sure
that if the number of unpunctured transition nodes is greater than
the number of information bit nodes, then any information bit is
connected to at least an unpunctured transition node.

Now that we have constructed the PEG-based interleaver, we can
derive a lower-bound of the girth of the corresponding factor graph.
Similar to the proof of Theorem 1 in \cite{PEG}, we can easily show
that:\vspace{-1mm}

\footnotesize
\begin{equation}
g \geq 2 \left(\left\lfloor\frac{\log \left[N(k+2)\left(1 -
\frac{1}{\dmax}\right) - N + 1\right]}{\log\left[(\dmax -
1)(k+1)\right]} -1\right\rfloor + 2\right)
\end{equation}
\normalsize
 Otherwise formulated, the girth of an irregular PEG
interleaved factor graph increases with the logarithm of the
interleaver size, which is optimal according to
(\ref{g_upper_bound}).

Over the BEC, the performance under iterative decoding is determined
by the minimum {\em stopping set} \cite{di2002fla}
\cite{rosnes_turbo_2007} of the factor graph of the code, the size
of which is upper-bounded by its minimum distance. As this minimum
stopping set is in general related to the girth of a factor graph
\cite{orlitsky2002ssa}, we would expect its size increases with the
girth. Moreover, the minimum distance of a turbo code grows
logarithmically with the interleaver size
\cite{breiling_logarithmic_2004}.  The rate of growth  of the
minimum stopping set of turbo codes under PEG-based interleaving
would thus be optimal.

\section{Simulation results}
\label{sim} 
\begin{figure}[h]
\begin{center}
   \includegraphics[width=1\columnwidth,angle=0]{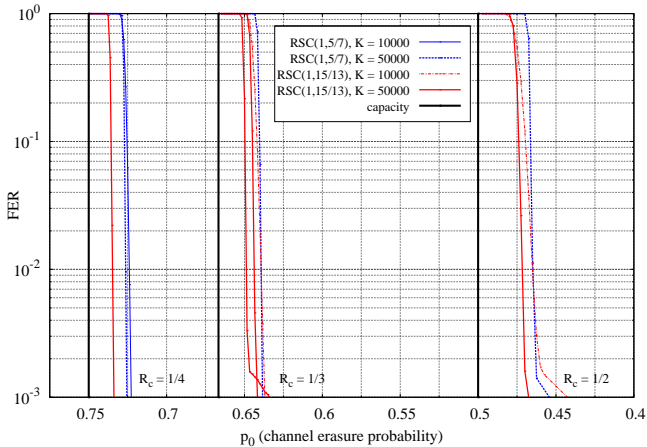}
   \vspace{-9mm}
   \caption{Performance of irregular turbo codes over the BEC.}
   \label{perf}
\end{center}
\end{figure}

In this section, frame error rate performance of
irregular turbo codes over the BEC is shown. Although the analysis
so far is based on the BCJR algorithm that supposes soft information
exchange, it was shown in \cite{kurkoski_analysis_2004} that a
hard-input hard-output (HIHO) decoding algorithm (namely the Viterbi
algorithm \cite{viterbi_error_1967}) for convolutional codes is
optimal in terms of bit error probability over the BEC. For this
reason, we will use a HIHO decoding algorithm for irregular turbo
codes inspired by the algorithm in \cite{luby1997plr} for LDPC
codes, in that it propagates in the trellis of the turbo code by
removing transitions in the same way edges are removed in a
bipartite graph under message-passing decoding \cite{kraidy-2008}.
This decoding scheme ensures a decoding complexity linear in the
interleaver size. Codes from Table \ref{table_1}, although having
very high thresholds, suffer from high error floors ($>10^{-2}$).
However, they are well suited for applications for which the quality
criterion is the average inefficiency \cite{kraidy-2008}. Fig.
\ref{perf} shows the performance of irregular turbo codes having a
good threshold-error floor tradeoff. In order to avoid high error
floors, the $8$-state RSC $(1,15/13)_8$ code was used as the
constituent code of the irregular turbo code. At a target frame
error rate of about $10^{-3}$, irregular turbo codes are within
$0.018 \leq \Delta_p \leq 0.028$ from capacity for various coding
rates.

\vspace{-5mm}
\section{Conclusions}
\label{conc} We proposed irregular turbo codes that perform close to
capacity for the binary erasure channel.  The codes operate for
various coding rates, and they provide low error floors with a
PEG-based interleaver that maximizes the cycles in the graphical
representation of the code. Implemented with an ``on-the-fly''
hard-input hard-output decoding algorithm, we believe that these
codes are suited for software implementation in upper-layer forward
error correction (UL-FEC) contexts.

\bibliographystyle{IEEEbib}
\footnotesize
\bibliography{strings,zotero}

\end{document}